\documentclass[letter,structabstract,longauth]{aa} 
 
\usepackage{graphicx}
\usepackage{txfonts}

\newcommand{\onvire}[1]{}

\newcommand{\beq}{\begin{equation}}
\newcommand{\eeq}{\end{equation}}

\usepackage{color}
\usepackage{epsfig}
\usepackage{amssymb}
\usepackage{graphicx,natbib}
\usepackage{pdflscape}
\usepackage{longtable}
\usepackage[figuresright]{rotating} 
\begin{document}

\title{Warm dust resolved in the cold disk around T\,Cha with
  VLTI/AMBER\thanks{Based on AMBER observations collected at the VLTI
    (European Southern Observatory, Paranal, Chile) with open time
    programs 083.C-0295(A,B).}}
    
   \author{
     J. Olofsson
     \inst{1}
     \and 
     M. Benisty
     \inst{1}
     \and 
     J.-C. Augereau
    \inst{2}
    \and 
    C. Pinte
    \inst{2} 
   \and
    F. M\'enard
    \inst{2}
    \and 
    E. Tatulli
    \inst{2}
    \and
    J.-P. Berger
    \inst{3}
    \and
    F. Malbet
    \inst{2}
    \and
    B. Mer\'{\i}n
    \inst{4}
    \and
    E.~F. van Dishoeck
    \inst{5,6}
    \and
    S. Lacour
    \inst{7}
    \and
    K. M. Pontoppidan
    \inst{8}
    \and
    J.-L. Monin
    \inst{2}
    \and 
    J. M. Brown
    \inst{6}
    \and
    G. A. Blake
    \inst{8}
         }

   \offprints{olofsson@mpia.de}
   
   \institute{Max Planck Institut f\"ur Astronomie,
     K\"onigstuhl 17, 69117 Heidelberg, Germany \\
     \email{olofsson@mpia.de} \and
     UJF-Grenoble 1 / CNRS-INSU, Institut de Plan\'etologie et
     d'Astrophysique de Grenoble (IPAG) UMR 5274, Grenoble, F-38041,
     France.\and
     ESO, Alonso de Córdova 3107, Vitacura, Casilla 19001, Santiago de
     Chile, Chile \and
     Herschel Science Centre, SRE-SDH, ESA P.O. Box 78, 28691
     Villanueva de la Ca\~nada, Madrid, Spain \and
    Leiden Observatory, Leiden University, P.O. Box 9513, 2300 RA
     Leiden, The Netherlands \and
     Max Planck Institut f\"ur Extraterrestrische Physik,
     Giessenbachstrasse 1, 85748 Garching, Germany \and 
     Observatoire de Paris, LESIA/CNRS (UMR 8109), 5 place Jules Janssen,
     Meudon, France \and
     California Institute of Technology, Division of Geological and
     Planetary Sciences, MS 150-21, Pasadena, CA 91125, USA
   }

   \date{Received \today; accepted }

   \abstract
   {The transition between massive Class~II circumstellar disks and
     Class~III debris disks, with dust residuals, has not yet been clearly
     understood. Disks are expected to dissipate with time, and dust
     clearing in the inner regions can be the consequence of several
     mechanisms. Planetary formation is one of them that will possibly
     open a gap inside the disk.
   }
   {According to recent models based on photometric observations,
     T\,Cha is expected to present a large gap within its disk,
     meaning that an inner dusty disk is supposed to have survived
     close to the star. We investigate this scenario with new
     near-infrared interferometric observations.  
   }
   {We observed T\,Cha in the $H$ and $K$ bands using the AMBER
     instrument at VLTI and used the MCFOST radiative transfer code to
     model the SED of T\,Cha and the interferometric
     observations simultaneously and to test the scenario of an inner dusty
     structure. We also used a toy model of a binary to check that a
     companion close to the star can reproduce our observations. }
   {The scenario of a close (few mas) companion cannot satisfactorily
     reproduce the visibilities and SED, while a disk model with a
     large gap and an inner ring producing the bulk of the emission
     (in $H$ and $K$-bands) close to 0.1\,AU is able to account for
     all the observations.  }
   {With this study, the presence of an optically thick inner dusty
     disk close to the star and dominating the $H$ and $K-$ bands
     emission is confirmed. According to our model, the large gap
     extends up to $\sim$\,7.5\,AU. This points toward a companion
     (located at several AU) gap-opening scenario to explain the
     morphology of T\,Cha.}
   \keywords{Stars: pre-main sequence -- evolution -- planetary
     systems: protoplanetary disks -- circumstellar matter --
     Infrared: stars -- Techniques: interferometric}
\authorrunning{Olofsson et al.}
\titlerunning{VLTI/AMBER observations of the T~Tauri star T\,Cha}

   \maketitle
%

\section{Introduction}

Using the {\it IRAS} space observatory, \citet{Strom1989} have reported
that several pre-main-sequence stars show a lack of emission in the
mid-infrared (IR) compared to their far-IR flux, owing to a lack of warm
dust indicative of the early stages of planetary formation. Several
Herbig Ae/Be stars also show very similar properties in their spectral
energy distribution (SED hereafter), such as \object{HD100546}
(\citealp{Bouwman2003}) observed with {\it ISO}. Thanks to the great
sensitivity of the {\it Spitzer Space Telescope}, fainter T~Tauri
stars were found to present such behavior (e.g.,
\citealp{Brown2007}). These so-called ``cold disks'' are suspected to
represent one of the transitional stages between the dust and gas-rich
Class~II objects and the Class~III stars with only dust and gas
residuals. The lack of emission in the mid-IR domain has been interpreted
as a tracer for dust clearing in the first AU of the circumstellar
disks. Several mechanisms can be responsible for dust clearing, such
as a binary companion (\citealp{Jensen1997}) which is the case for
CoKu Tau 4 (\citealp{Ireland2008}), or photo-evaporation processes
(\citealp{Alexander2006}). These mechanisms will clear the disk 
as probed by submillimeter and millimeter observations (e.g.,
\citealp{Andrews2009}, \citealp{Hughes2009},
\citealp{Brown2009}). Another dust-clearing process is the formation
of a planet inside the disk. Because the planet is accreting surrounding
material, a gap may open within the disk, resulting in an inner and an
outer disks (e.g., \citealp{Varni`ere2006}).

In this Letter, we present the first near-IR interferometric
observations of a T~Tauri star by the AMBER instrument, installed at
the Very Large Telescope Interferometer
\citep[VLTI;][]{Scholler2007}. \object{T\,Cha} (spectral type G8) is
one of the four cold disks studied by \citet{Brown2007}. To
model the SED of T\,Cha, and especially the Spitzer/IRS spectrum, they
needed two separated regions in the disk: an inner low-mass radially
confined belt (from 0.08 to 0.2\,AU) and a massive outer disk
(15--300\,AU). The first component accounts for the excess emission at
wavelengths shorter than 10\,$\mu$m, while the second accounts for the
emission at longer wavelengths. The large gap allows the lack of
emission around 10\,$\mu$m to be reproduced. As the presence of a gap
may suggest there are substellar companions, T\,Cha would
therefore become a prime target when searching for planets. Unfortunately,
SED fitting is highly degenerated, therefore we acquired spatially
resolved observations at high angular resolution with the near-IR
instrument, AMBER, at the VLTI to lift some of the degeneracies in the
SED fitting process, to confirm the presence of the inner disk, and
study its structure.

We first describe, in Sect.\,\ref{sec:obs}, the AMBER observations and
their reduction, as well as first simple models used to interpret
these data. In Sect.\,\ref{sec:mod} we present radiative transfer
models that we use to reproduce the SED and
interferometric observables simultaneously, and we finally conclude with the results in
Sect.\,\ref{sec:disc}.

\section{Results and first analysis\label{sec:obs}}
\subsection{VLTI/AMBER observations}
\vspace*{-0.1cm}
\begin{figure*}
\begin{center}
\hspace*{-0.cm}\includegraphics[angle=0,height=0.57\columnwidth]{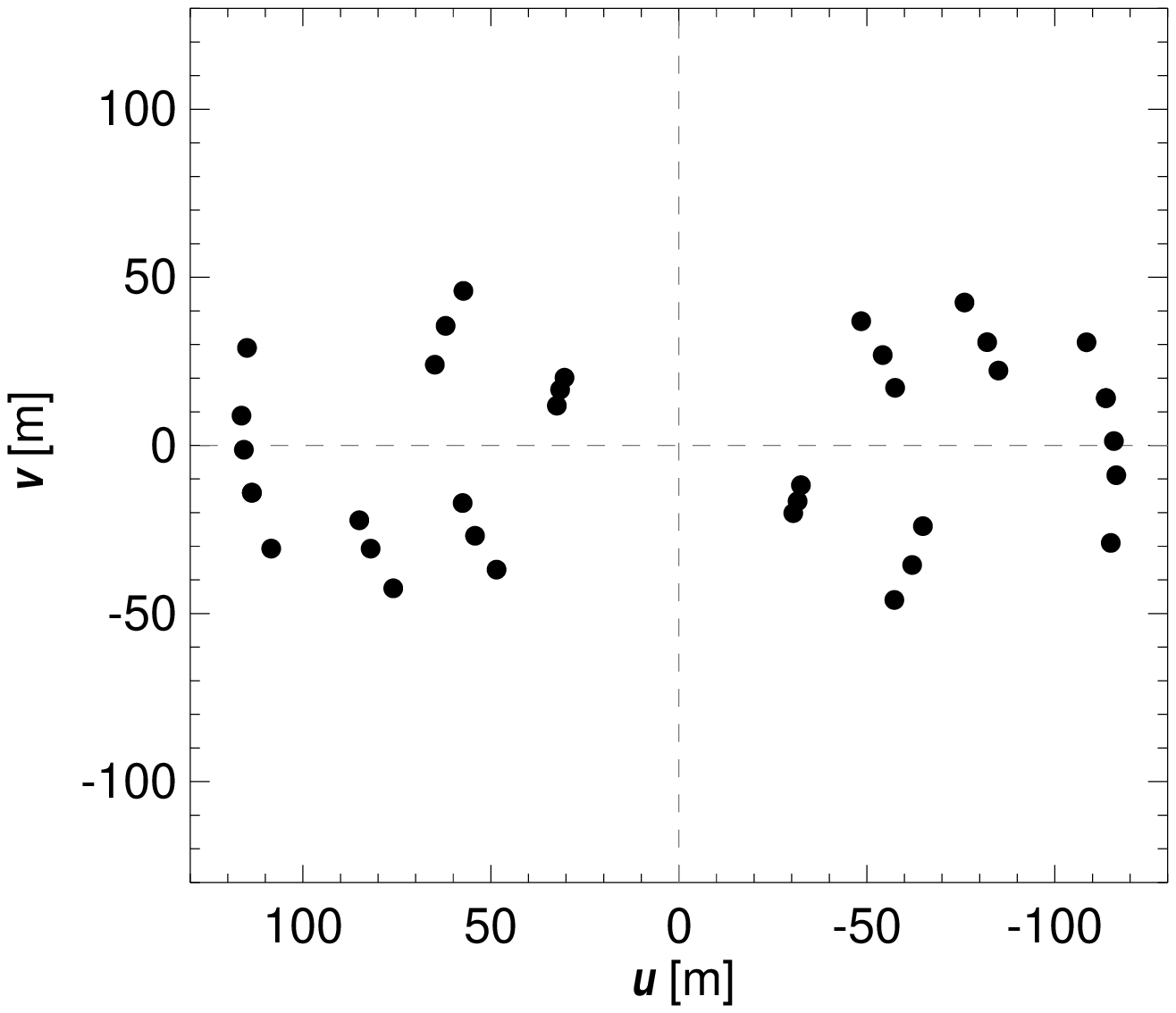}
\hspace*{-0.cm}\includegraphics[angle=0,height=0.57\columnwidth]{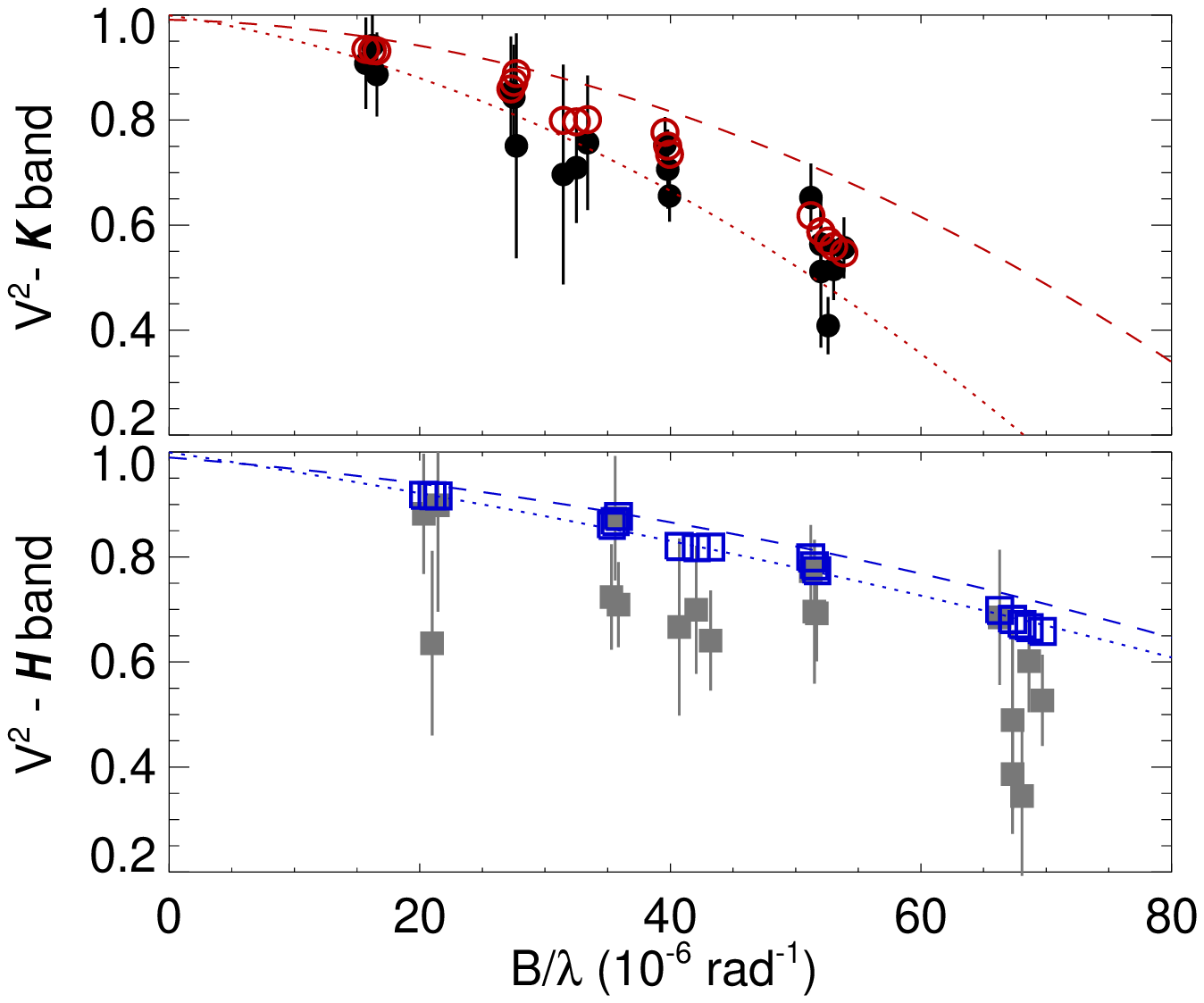}
\hspace*{-0.cm}\includegraphics[angle=0,height=0.57\columnwidth]{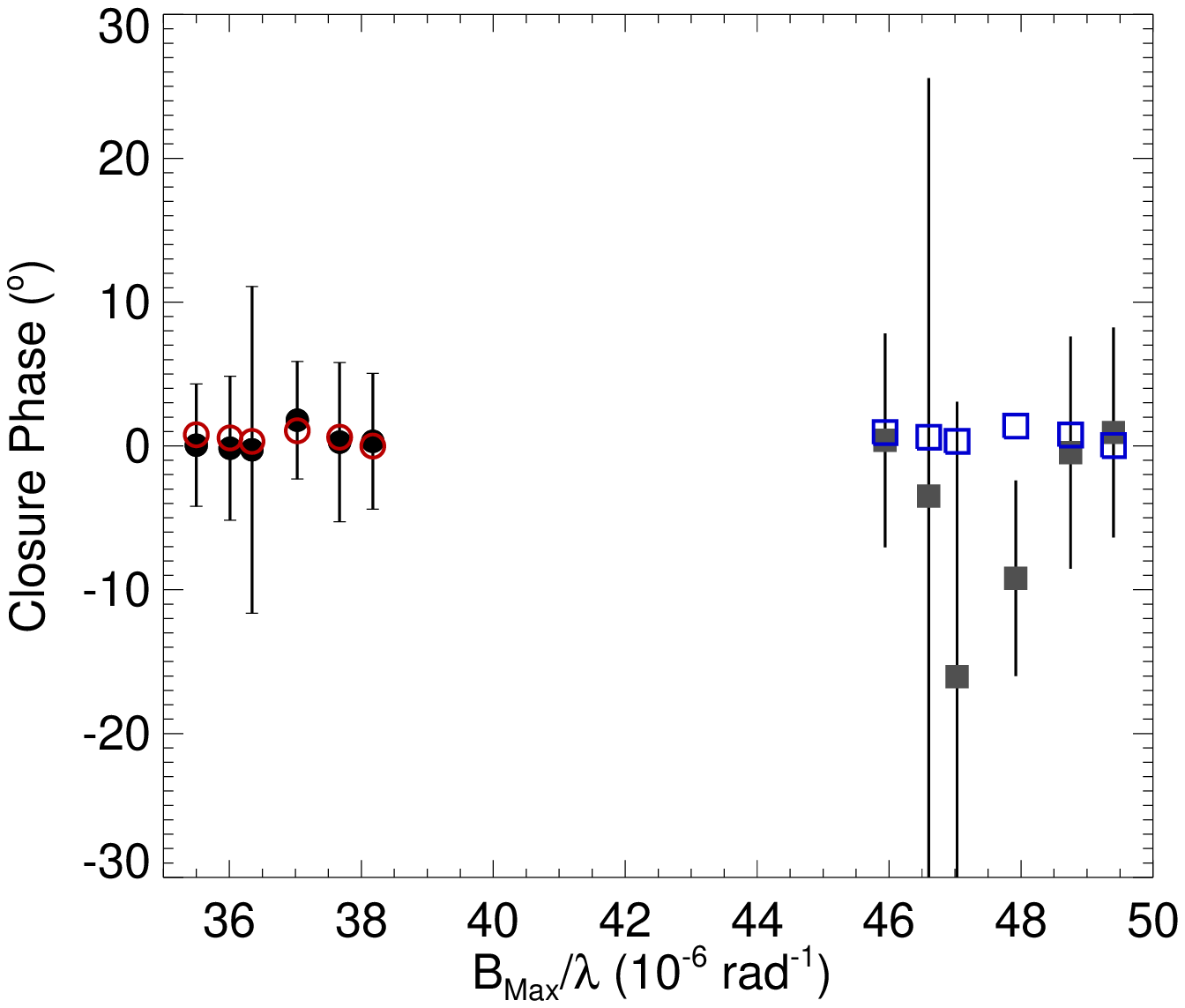}
\vspace*{-0.3cm}
\caption{{\it Left panel}: $(u,v)$ coverage of the
  observations. Observed squared visibilities ({\it middle}) and CP
  ({\it right}). $H$ and $K$-bands data are represented by squares and
  circles symbols, respectively. Overplotted are the predictions from
  our best disk model (empty symbols). Dashed and dotted lines in the
  middle panels are a $2^{\mathrm{nd}}$-order polynomial fit the modeled 
  visibilities for the extreme values on R$_{\mathrm{in,\,inner\,disk}}$ 
  within the uncertainties (see
  Sect.\,\ref{sec:mod}).\label{fig:vis}}
\end{center}
\vspace*{-0.5cm}
\end{figure*}
T\,Cha was observed at the VLTI, using the AMBER instrument that
allows the simultaneous combination of three beams in the near-IR with
spatial filtering (\citealp{Petrov2007}).  The instrument delivers
spectrally dispersed interferometric visibilities, closure phases (CP
hereafter), and differential phases at spectral resolutions up to
12~000. In the following, we present $K$- and $H$-bands (around 2.2
and 1.65\,$\mu$m, respectively) observations taken in the low spectral
resolution mode (LR; $R \sim$35) with the 8.2~m Unit Telescopes (UTs),
coupled with the use of adaptive optics.  T\,Cha was observed with 5
different baselines of two VLTI configurations (UT1-2-4 and UT1-3-4),
during June 11 and 12, 2009.  The longest baseline length is
$\sim$130\,m corresponding to a maximum angular resolution of
1.3\,milli-arcseconds (mas). The integration time per frame was
50\,ms. In addition to T\,Cha, a calibrator (HD\,107145, F5V) was
observed to correct for instrumental effects.  Observations were
performed without fringe-tracking.

Data was reduced following the standard procedures
described in \citet{Tatulli2007} and \citet{Chelli2009}, using the
\texttt{amdlib} package, release 2.99, and the \texttt{yorick}
interface provided by the Jean-Marie Mariotti
Center ({$\textrm{http://www.jmmc.fr}$}). Raw spectral visibilities and CP were extracted for all the frames of
each observing file.  A selection of 80\% of the highest quality
frames was made to minimize the effect of instrumental jitter and
non-optimal light injection into the instrument. The
AMBER+VLTI instrumental transfer function was calibrated using measurements
of \object{HD\,107145}, after correcting for its diameter ($0.3 \pm
0.1$\,mas, which was obtained using the getCal software,
$\textrm{http://nexsciweb.ipac.caltech.edu/gcWeb/gcWeb.jsp}$). The $H$-band
measurements have a lower signal-to-noise ratio since they have been
obtained close to the instrument-limiting magnitude. With about 20
spectral channels through the $H$ and $K$~bands, the total data set
(after processing) is made of 585 and 195 spectrally dispersed
visibilities and CP, respectively (averaged for each baseline and
baseline triplet to provide 36 and 12 broad-band visibilities and CP,
respectively).

Figure\,\ref{fig:vis} presents the final reduced data for
\object{T\,Cha}: the $(u,v)$ coverage (left panel), the squared
visibilities in $H$ and $K$ broad bands as a function of the spatial
frequencies (gray and black, respectively, middle panel), and the CP
(right panel). The squared visibilities show that the circumstellar
emission around T\,Cha is marginally resolved: the squared
visibilities decrease with increasing spatial frequency, but
remain at a high level (V$^2 \ge 0.3$). We do not observe strong
discrepancies between $H$ and $K$ bands visibilities, suggesting that
the same structure is responsible for most of the emission in $H$ and
$K$ bands. Finally, the CP are close to zero, indicative of a
centro-symmetric structure at the given spatial resolution of the
observations. We first investigate the possibility of a companion
around T\,Cha, and then examine whether an inner dusty disk as proposed by
\citet{Brown2007} can reproduce the data.  \vspace*{-0.4cm}

\subsection{Geometric models\label{sec:toy}}
\vspace*{-0.1cm}

To see whether our interferometric datapoints are compatible with
a companion around T\,Cha, we used a toy model that aims to describe a
binary system where both stars are individually unresolved (e.g.,
\citealp{Lawson1999}). To find a model that best reproduces both AMBER
visibilities and CP, we performed a $\chi^2$ minimization exploration
through a 3-dimensional grid: flux ratio between the two components
($F_2/F_1$), separation ($\rho$), and position angle ($PA$). The
reduced $\chi^2_{\mathrm{r}}$ values are then transformed into
normalized probabilities to perform a statistical (Bayesian)
analysis. The top three panels of Fig.\,\ref{fig:bin} display the
reduced $\chi^2_{\mathrm{r}}$ maps for the three pairs of free
parameters. We derived uncertainties from the probability
distributions. The binary parameters for the best model
($\chi^2_{\mathrm{r}} =$ 0.29) are the following: flux ratio of
$F_2/F_1 = 0.95_{-0.85}^{+0.05}$, separation of $\rho =
1.15_{-0.49}^{+0.57}$\,mas, and position angle on the sky of $PA =
68^{\circ}$$^{+25} _{-68}$ east of north. The uncertainty on the flux
ratio is rather large, because the squared visibilities (36 values)
can also be reproduced for a small-flux ratio. However, in that
case, CP (12 values) are not close to zero and CP are only matched for
higher flux ratio, close to unity. The large uncertainties on
$F_2/F_1$ therefore reflect the larger amount of visibility with
respect to the number of CP. That the flux ratio is about
$F_2/F_1 \sim$\,1 implies that the emission profile of the companion,
if any, would be similar to the one of the primary and not colder, as
expected in order to match the near-IR excess probed from SED modeling
(\citealp{Brown2007}). Besides this consideration, \citet{Schisano2009}
conclude from radial velocity measurements, that even a low-mass
object (0.05\,$M_{\odot}$) at a distance of 0.1\,AU would have been
detected. Additionally, NaCo narrow-band imaging observations of
T\,Cha at 2.12 and 1.75\,$\mu$m rule out any stellar companion
with a mass higher than 0.15\,$M_{\odot}$ between $\sim$2 and 10\,AU
(Vicente et al. submitted). A close companion in the first tenths of
AU around T\,Cha is therefore not a satisfying solution.

We now consider the case of a uniform ring aimed at representing the
inner disk edge (e.g., \citealp{Eisner2004}). The three input
variables of this model are the position angle of the disk ($PA$,
$0^{\circ}$, meaning the major semi-axis is oriented north-south), the
inclination on the line of sight ($i$, $0^{\circ}$ is face-on), and
the inner diameter ($\theta$). We fixed the star-to-ring flux ratio to
account for $H$ and $K$ band excess derived using a Nextgen synthetic
spectrum of the star (see Sect.\,\ref{sec:mcfost}). The width $W$ of
the uniform ring is given by $W/( \theta /2) = 0.18$. We performed a
grid exploration for these three parameters, with a
$\chi^2_{\mathrm{r}}$ minimization. The bottom three panels of
Fig.\,\ref{fig:bin} show the reduced $\chi^2_{\mathrm{r}}$ maps for
the three pairs of free parameters. Only the diameter is
determined well with this approach, while the position angle and
inclination cannot be strongly constrained, because of the almost
unidirectional alignment of the $(u, v)$ points. The final best fit
($\chi^2_{\mathrm{r}}$ = 0.76) returns $PA =
67^{\circ}$$_{-56}^{+91}$, $i = 42^{\circ}$$_{-40}^{+32}$, and $\theta
= 1.82_{-0.38}^{+0.52}$\,mas. Therefore, a dust ring model at a tenth
of AU or so from the star is consistent with the AMBER visibilities
and with the photometric excesses in the $H$- and
$K$-bands. Nonetheless, the above model is an oversimplified
description of a disk and ignores the disk's asymmetry, as well as the
SED at other wavelengths. We develop a refined model in the next
section.

\vspace*{-0.3cm}
\section{Radiative transfer modeling\label{sec:mod}}

Hipparcos has estimated a $66^{+19}_{-12}$\,pc distance for T\,Cha,
but, as explained in \citet{Schisano2009}, proper motions measurements
by \citet{Frink1998} and \citet{Terranegra1999} suggest that this
distance is underestimated. T\,Cha may in fact belong to a star
association cluster located at 100\,pc, which is the distance that we
adopt in this study.
The disk inclination is also uncertain. An
inclination of 75$^{\circ}$ is commonly used in the literature. This value comes from
$v$\,sin\,$i$ measurements (54\,km.s$^{-1}$, \citealp{Alcal'a1993})
indicative of a highly inclined object. \citet{Schisano2009} measured
$v$\,sin\,$i$ of 37\,$\pm$2\,km.s$^{-1}$, which could suggest a
slightly smaller inclination. However, T\,Cha displays CO in
absorption in the 4.7\,$\mu$m fundamental band (Brown et al. in prep),
which requires a high inclination ($\sim$60--70$^{\circ}$). In the
following, the inclination is therefore considered as a free parameter
and is further discussed in Sect.\,\ref{sec:mcfost}.
\vspace*{-0.2cm}

\subsection{Model setup}
\vspace*{-0.1cm}
To model the disk emission and to reproduce simultaneously, the
squared visibilities and CP in $H$ \& $K$ bands, as well as the SED,
we used the MCFOST radiative transfer code
(\citealp{Pinte2006}). MCFOST calculates synthetic observables, such
as SEDs and monochromatic raytraced images. The computation is done by
propagating packets of photons inside the disk. Light scattering,
absorption, and re-emission processes are included. The geometry of
the disk is defined by several parameters: the inner and outer
radii ($R_{\mathrm{in}}$ and $R_{\mathrm{out}}$, respectively), the
index $\alpha$ for the surface density ($\Sigma (r) = \Sigma
(r/r_0)^{\alpha}$), and a disk scale height H(r) assuming a vertical
Gaussian distribution ($\exp[-z^2/2H(r)^2]$), germane to a disk at the
hydrostatic equilibrium. The disk's flaring is described by a
powerlaw determining the scale height as a function of the radius
($H(r) = H_0(r/r_0)^{\beta}$). The dust content is described by a
differential powerlaw for the grain size distribution (d$n(a) \propto
a^{-p}$d$a$), between the minimum ($a_{\mathrm{min}}$) and maximal
($a_{\mathrm{max}}$) grain sizes. The radiative transfer code can
handle spatially separated disk zones.
\begin{figure}
\begin{center}
\includegraphics[angle=0,width=1.\columnwidth,origin=bl]{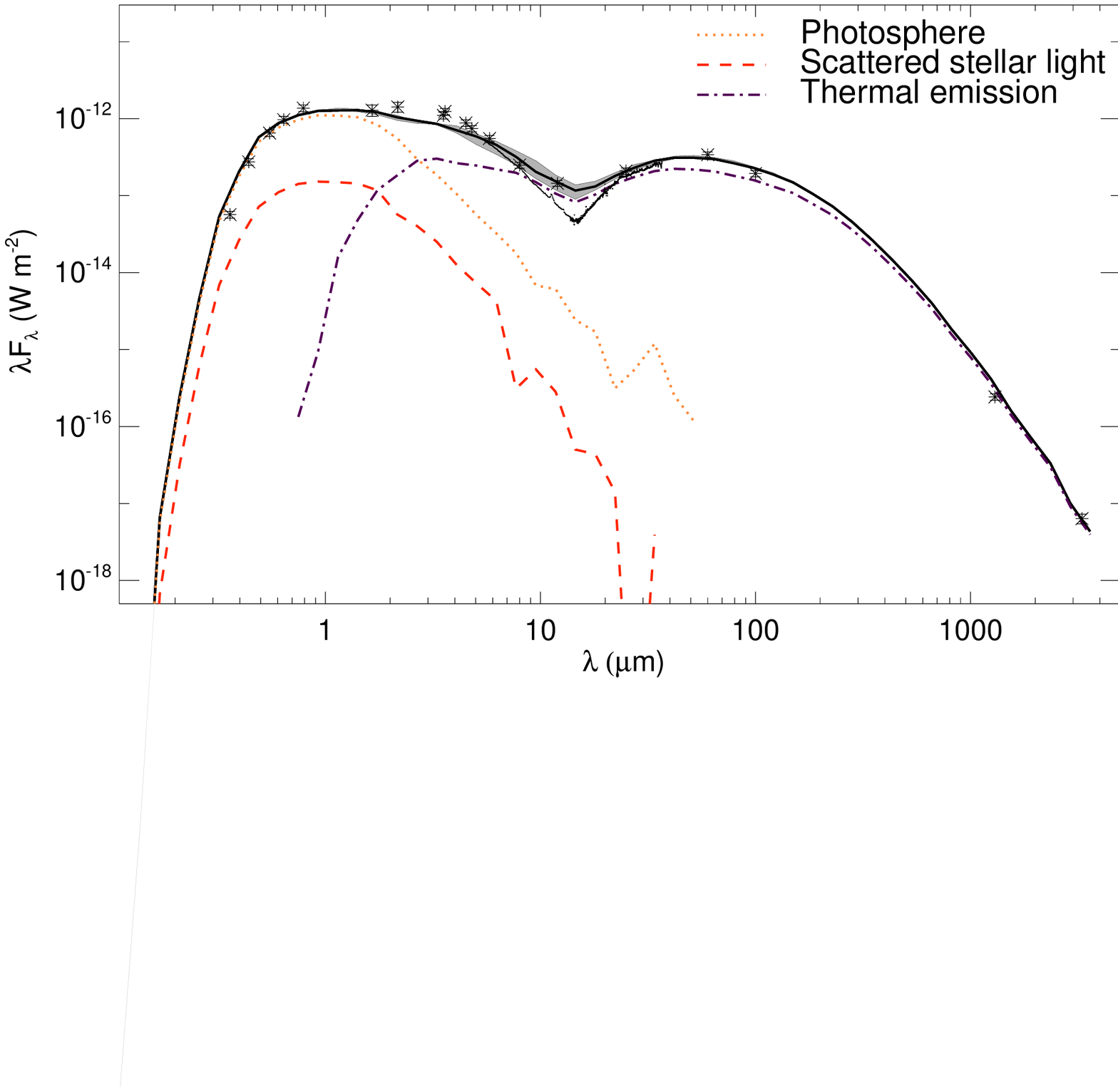}
\vspace*{-0.4cm}
\caption{ Observed and modeled SED of T\,Cha. Photometric measurements
  were gathered from the literature (\citealp{Brown2007},
  \citealp{Lommen2010}, \citealp{Alcal'a1993} and reference
  therein). The gray area represents the modeled SED for the extreme
  values on R$_{\mathrm{in,\,inner\,disk}}$, within the uncertainties
  (see Sect.\,\ref{sec:mod}).\label{fig:sed}}
\end{center}
\vspace*{-0.5cm}
\end{figure}
\begin{table}
\centering
\caption{Parameters of the disk model} 
\label{tab:mcfost}
\begin{tabular}{l|cc}
  \hline
  Parameter & Inner disk & Outer disk  \\ 
  \hline
  Size distribution index $p$ & -3.5 & -3.5 \\
  a$_{\mathrm{min}}$ -- a$_{\mathrm{max}}$ [$\mu$m] & 0.1 -- 10 & 0.01
  -- 3000 \\
  M$_{\mathrm{dust, disk}}$ [M$_{\odot}$] & 1\,$\times 10^{-9}$ & 1.75\,$\pm 0.25 \times 10^{-4}$ \\
  R$_{\mathrm{in}}$ [AU] & 0.13$\pm 0.03$ & 7.5$\pm 0.5$ \\
  R$_{\mathrm{out}}$ [AU] & 0.17$\pm 0.03$ & 300 \\
  Flaring index ($\beta$) & 1.11 & 1.11 \\
  $H/R$ (at 1\,AU) & 0.106 & 0.15 \\
  Surface density index ($\alpha$) & -1 & -1 \\
  \hline
\end{tabular}
\vspace*{-0.3cm}
\end{table}
\vspace*{-0.2cm}

\subsection{Results and discussion\label{sec:mcfost}}
\vspace*{-0.1cm}

To reproduce the stellar photosphere, we adopt a NextGen stellar
atmosphere model (\citealp{Allard1997}). All the stellar parameters,
except the inclination $i$, were taken from \citet{Brown2007}. The two
parameters $T_{\mathrm{eff}}$ and $R_{\ast}$ were then varied only to
remain unchanged once the $UBVRI$ bands were reproduced well enough. In the
end, the star is defined with the following parameters:
$T_{\mathrm{eff}} = 5400$\,K, log($g$) $= 3.5$, $R_{\ast} =
1.3\,R_{\odot}$, $M_{\ast} = 1.5\,M_{\odot}$, and $A_{\mathrm{v}} =
1.5$.

The disk is known to have a depleted zone (gap) according to SED modeling
(\citealp{Brown2007}). Therefore, two separate zones were defined for
the circumstellar material: an inner and an outer dusty
structure. Given the high angular resolution of our observations, the
outer disk is overresolved, so changing its characteristics will only
affect the SED. For the inner disk we use a mixture of astro silicates
(\citealp{Draine1984}) and amorphous carbon (\citealp{Zubko1996}) with
a mass ratio of 4:1 between silicate and carbon. We choose to use this
composition, because these grains will produce a very weak emission
feature at 10\,$\mu$m, which has not been detected in the IRS spectrum
(\citealp{Brown2007}). From previous tests with different
compositions (e.g., less carbon), such features, even a small one,
will fill the dip in the 7-15\,$\mu$m range and will not produce a
good fit to the SED. We try using larger grains ($\geq 10$\,$\mu$m) to
avoid any emission features, without any satisfying results on both
visibilities and SED. In our model,the dust mass of the inner disk is
10$^{-9}$\,M$_{\odot}$, and falls in the optically thick regime.  The
near-IR excess cannot be reproduced by an optically thin inner
disk. However, as the AMBER observations only trace the $\tau = 1$
surface, further constraints on the inner disk mass cannot be
established. For the outer disk, we use grain opacities of astro
silicates, and the total dust mass of the outer disk, for grain sizes
up to 3\,mm, of $[1.75 \pm 0.25] \times 10^{-4}$\,M$_{\odot}$ is
constrained well by the millimeter observations. We choose to use an
inclination of 60$^{\circ}$, while 75$^{\circ}$ is often used in the
literature. With an inclination higher than 60$^{\circ}$, the outer
disk hides the inner disk, which would account neither for the NIR excess
nor for the interferometric measurements. Even if the $PA$ does not
affect the SED, it affects the modeled visibilities and CP. It is
therefore important to estimate a reasonable value. We use the uniform
ring model described in Sect.\,\ref{sec:toy}, with a fixed inclination
of 60$^{\circ}$. The best $PA$ is in that case
78$^{\circ}$$^{+87}_{-50}$, and even if not well-constrained, we use
this value when computing visibilities from ray-traced images.

The best model ($\chi^2_{\mathrm{r}} = 0.58$) is found using a grid of
models on the following parameters: R$_{\mathrm{in}}$,
R$_{\mathrm{out}}$, M$_{\mathrm{dust}}$, and $H/R$ for the inner disk,
starting from the model described in \citet{Brown2007}. Similar models
are then computed for different grain compositions and sizes. Once the
visibilities and near-IR excess are reproduced well, we fine-tune the
outer disk's setup in order to match the far-IR and mm
observations. All the parameters of the disk model are compiled in
Table\,\ref{tab:mcfost}. The visibilities and CP for the best model,
calculated for the same ($u$, $v$) points (left panel of
Fig.\,\ref{fig:vis}) as the observations, are presented on the middle
and right panels of Fig.\,\ref{fig:vis}, respectively and the final
SED is displayed in Fig.\,\ref{fig:sed}. Modeled visibilities are
slightly overestimated compared to observations, especially in the
$H$-band, but lie within the uncertainties. Our
model reproduces well the overall decrease of the visibilities. 
Middle panels of Fig.\,\ref{fig:vis} also show the influence of the location 
of the inner ring models for the two extreme cases of R$_{\mathrm{in,\, inner\,disk}} = 0.1$ and
$0.16$\,AU, respectively, with R$_{\mathrm{out,\, inner\,disk}} =
$R$_{\mathrm{in,\, inner\,disk}} + 0.04$\,AU. The CP and SED (except
for the abrupt dip around 10\,$\mu$m) are also reproduced well by our 
two separate zones radiative transfer model. The
influence of the variation of R$_{\mathrm{in,\, inner\,disk}}$ in the
range $0.13 \pm 0.03$\,AU is also displayed on Fig.\,\ref{fig:sed}. 
This shows that, even with a large uncertainty on the $PA$ and a 
highly inclined disk, combining both
visibilities and SED provides strong constraints on the
R$_{\mathrm{in,\, inner\,disk}}$ parameter.

By combining the AMBER data and the SED, we constrain the inner dust
ring to be even narrower (between 0.13 and 0.17\,AU) than what was
inferred by \citet[][between 0.08 and 0.2\,AU]{Brown2007}. One should
note that our solution may not be unique, since material can possibly be
hidden at larger distances (few AU), shadowed by the inner disk, but
neither interferometric nor SED measurements can reveal such an
extension. Nevertheless, the location of the inner disk edge is
interestingly found to be almost similar when using MCFOST that
includes scattering and a geometric ring model that represents only
for the thermal emission ($R_{\mathrm{in}} = \theta /2 \sim 0.9$\,mas
$ = 0.09$\,AU at 100\,pc). This conclusion, in apparent contradiction
to \citet{Pinte2008a}, results from the extreme compactness of the
inner disk (width $\sim$\,0.04\,AU) that leads to a total emission for
which we cannot disentangle the respective contributions of thermal
and scattered light emissions. With the inner edge of the inner disk
set at 0.13\,AU, the dust reaches a maximum temperature of 1337\,K, in
agreement with carbon and silicate sublimation temperatures. Even if
the outer disk is not resolved with our observations, changing the
inner disk's setup (position, scale height, and flaring) has a strong
impact on the outer structure, because of the shadow cast by the inner
disk. We therefore had to shift the inner edge of the outer disk from
15\,AU (\citealp{Brown2007}) down to 7.5\,AU, in order to reproduce
the SED. Finally, we do not claim to have a precise determination of
either $i$ or $PA$ with our observations, considering the poor
($u, v$) coverage.

\section{Conclusion\label{sec:disc}}

In Sects.\,\ref{sec:obs} and \ref{sec:mod} we presented and modeled the
interferometric observations of T\,Cha obtained with the AMBER
instrument. Two models were investigated, and
using both near-IR visibilities and SED, we conclude that the scenario
of a close companion, in the first tenths of AU, cannot satisfactorily
reproduce the observations, while we present a successful radiative
transfer disk model that agrees with both interferometric data
and SED. With the near-IR visibilities, we spatially resolve and
witness the presence of warm dust and we constrain its location
(R$_{\mathrm{in,\,inner\,disk}}$). The thinness of the ring is derived
from combining the interferometric data and the SED, while the
gap is deduced solely from the SED. A continuous disk
model, without any gap and with an inner rim located at $0.13$\,AU would
also reproduce the near-IR visibilities, CP, and near-IR fluxes
($\chi^2_{\mathrm{r}} = 0.54$), but not the overall SED.

Models of gap opening by a Jupiter-mass planet located at 5\,AU (e.g.,
\citealp{Varni`ere2006}) predict a gap between $\sim$\,2 and 8\,AU,
suggesting either a multiple planet system, a more massive perturber, 
or an extended self-shadowed disk up to
$\sim$\,2\,AU. Interestingly, \citet{Hu'elamo2011} detected a $\sim$25\,M$_{\mathrm{Jup}}$ companion located at 6.7\,AU from T\,Cha using sparse aperture masking observations.
Future numerical simulations will have to (i) determine if
the inner disk can be a consequence of and survive the disk
clearing by possible companion(s) and (ii) investigate what the
contribution of inward mass transfer is throughout the gap to eventually
refill the inner disk.  If ignoring erosion mechanisms other than
accretion and assuming that no self-shadowed disk is present, the
survival of the disk over a few Myr can be ensured if the accretion rate
lies below $10^{-13}$\,M$_{\odot}.$yr$^{-1}$ (assuming a gas-to-dust
ratio of 100), a value consistent with the upper limit measured by
\citet{Schisano2009} to be of about $4 \times 10^{-9}$
M$_{\odot}.$yr$^{-1}$. Alternatively, we may be witnessing the
aftermath of a collision between two planetesimals or planet embryos,
resulting in a transient 2.7 M$_{\mathrm{Lunar}}$ dust belt around
T\,Cha.
\begin{acknowledgements}
  The authors thank the anonymous referee for the constructive and
  useful comments provided. They also thank the Max Planck Society,
  the Programme National de Physique Stellaire (PNPS) and ANR
  (contract ANR-07-BLAN-0221) for supporting part of this
  research. 
  C. Pinte acknowledges funding from the European Commission's
  7$^{\mathrm{th}}$ Framework Program (contracts PIEF-GA-2008-220891
  and PERG06-GA-2009-256513).
\end{acknowledgements}
\vspace*{-.7cm}

\bibliography{biblio}

\appendix
\section{Geometric models}

\begin{figure}[h]
\begin{center}
\hspace*{-0.5cm}\includegraphics[angle=0,height=0.32\columnwidth]{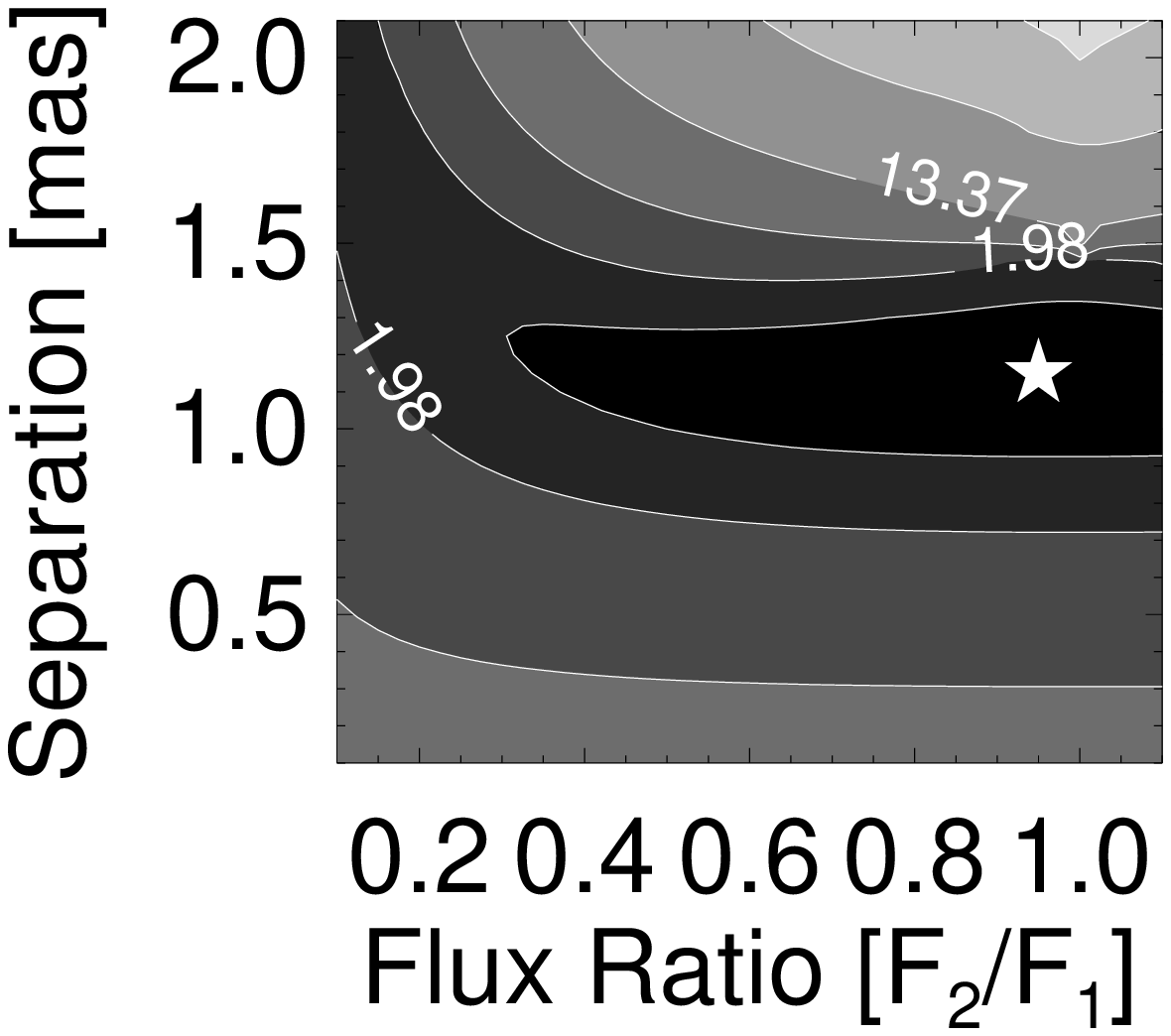}
\hspace*{-0.5cm}\includegraphics[angle=0,height=0.32\columnwidth]{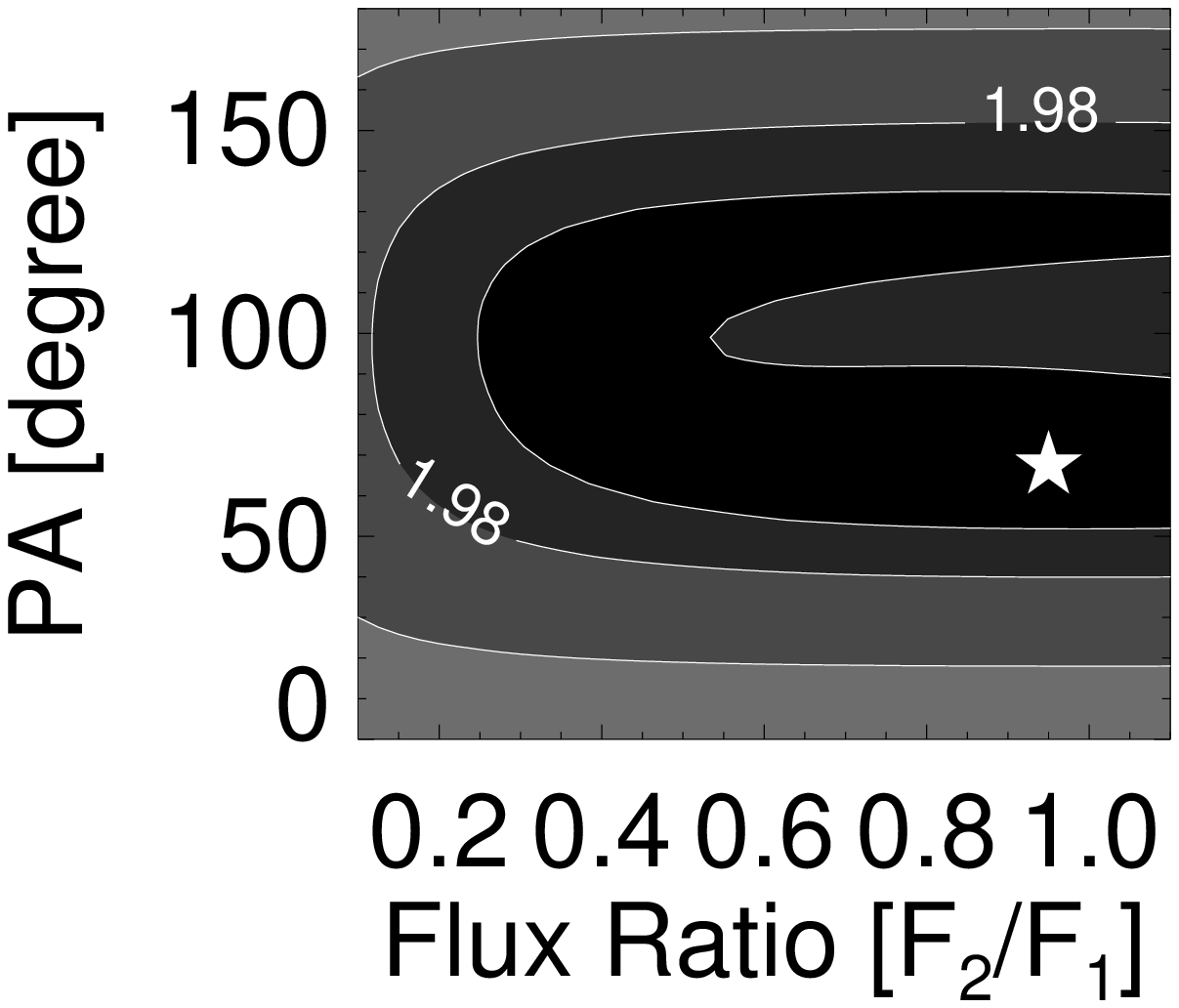}
\hspace*{-0.5cm}\includegraphics[angle=0,height=0.32\columnwidth]{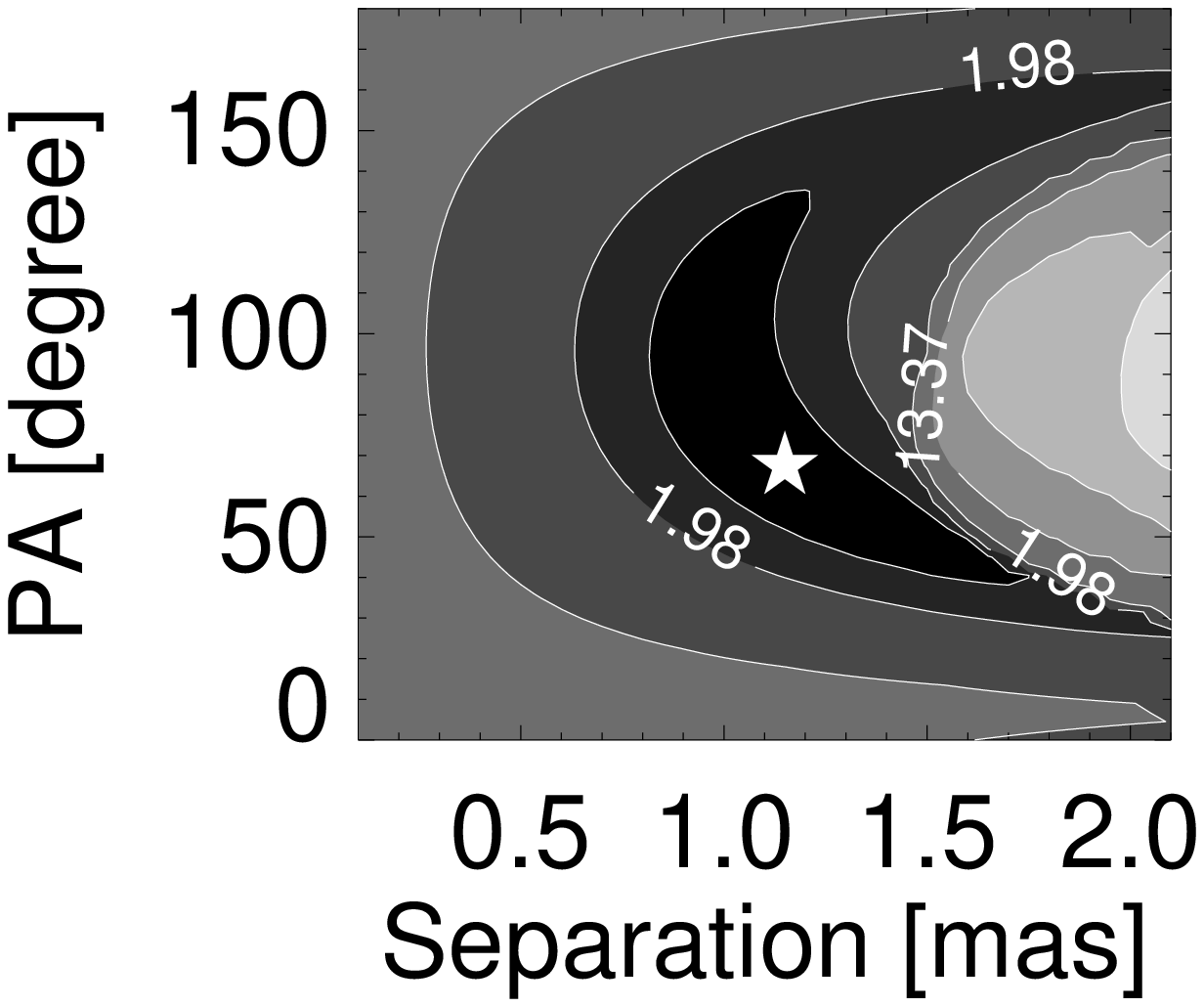}
\hspace*{-0.5cm}\includegraphics[angle=0,height=0.32\columnwidth]{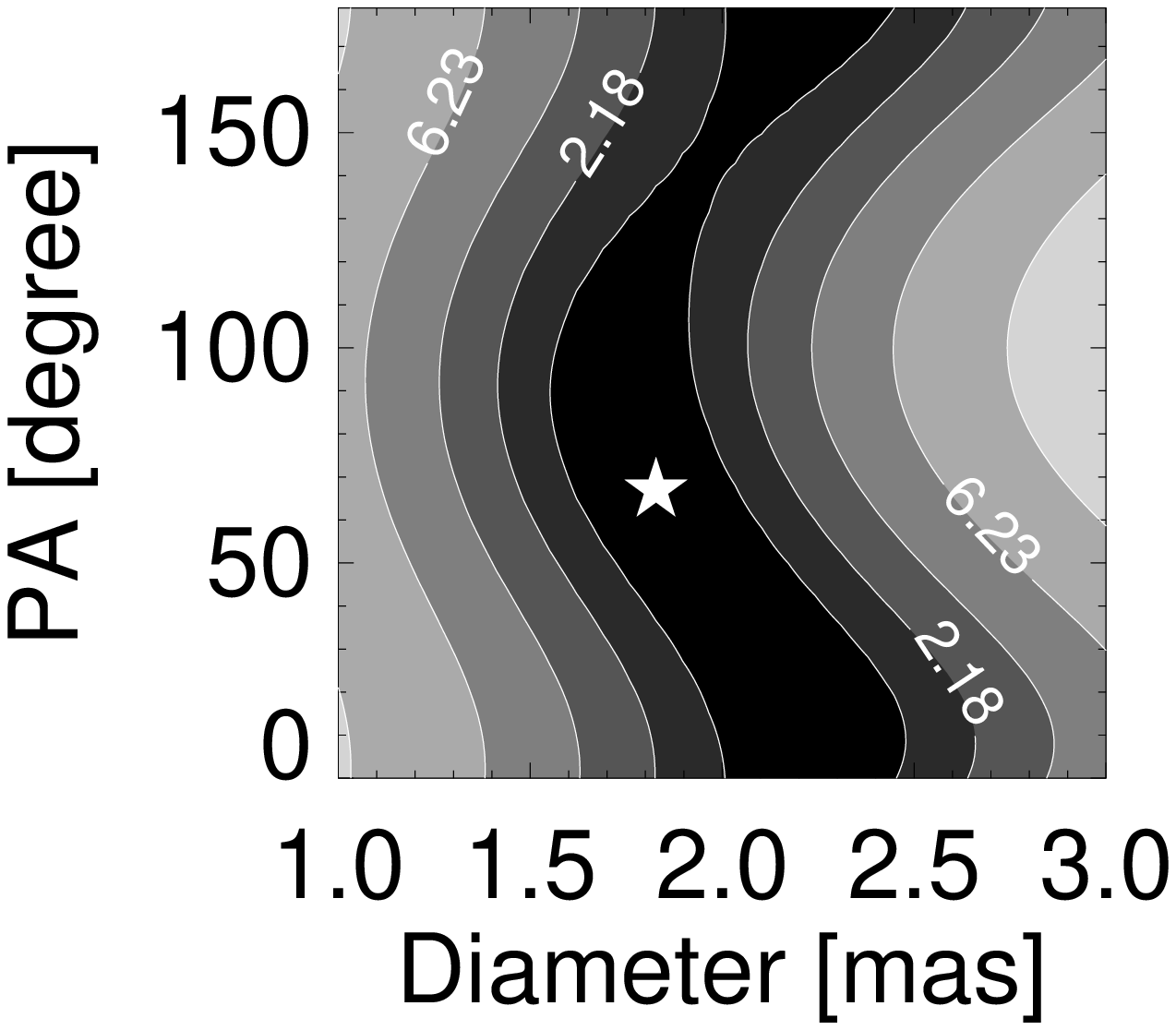}
\hspace*{-0.5cm}\includegraphics[angle=0,height=0.32\columnwidth]{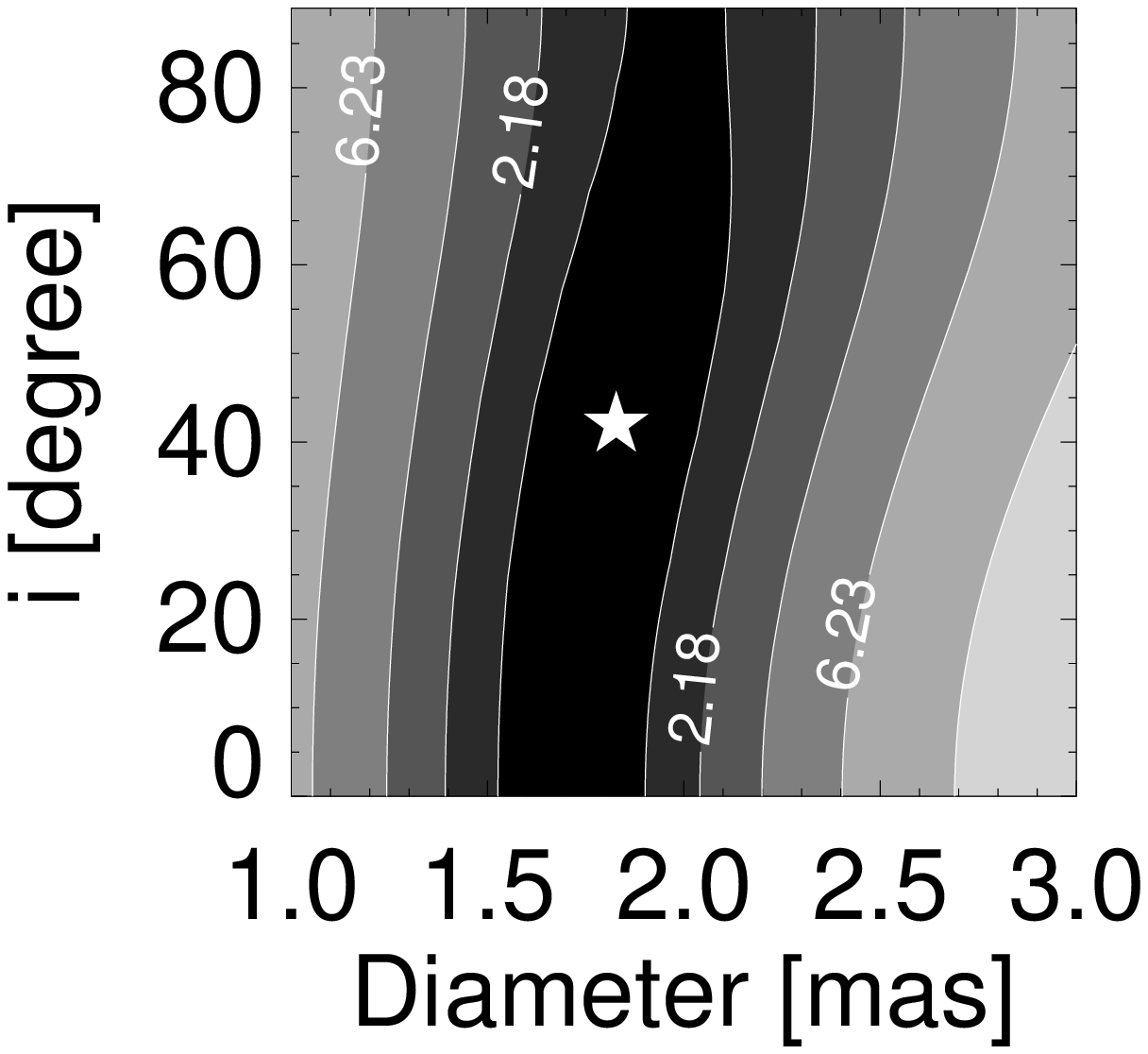}
\hspace*{-0.5cm}\includegraphics[angle=0,height=0.32\columnwidth]{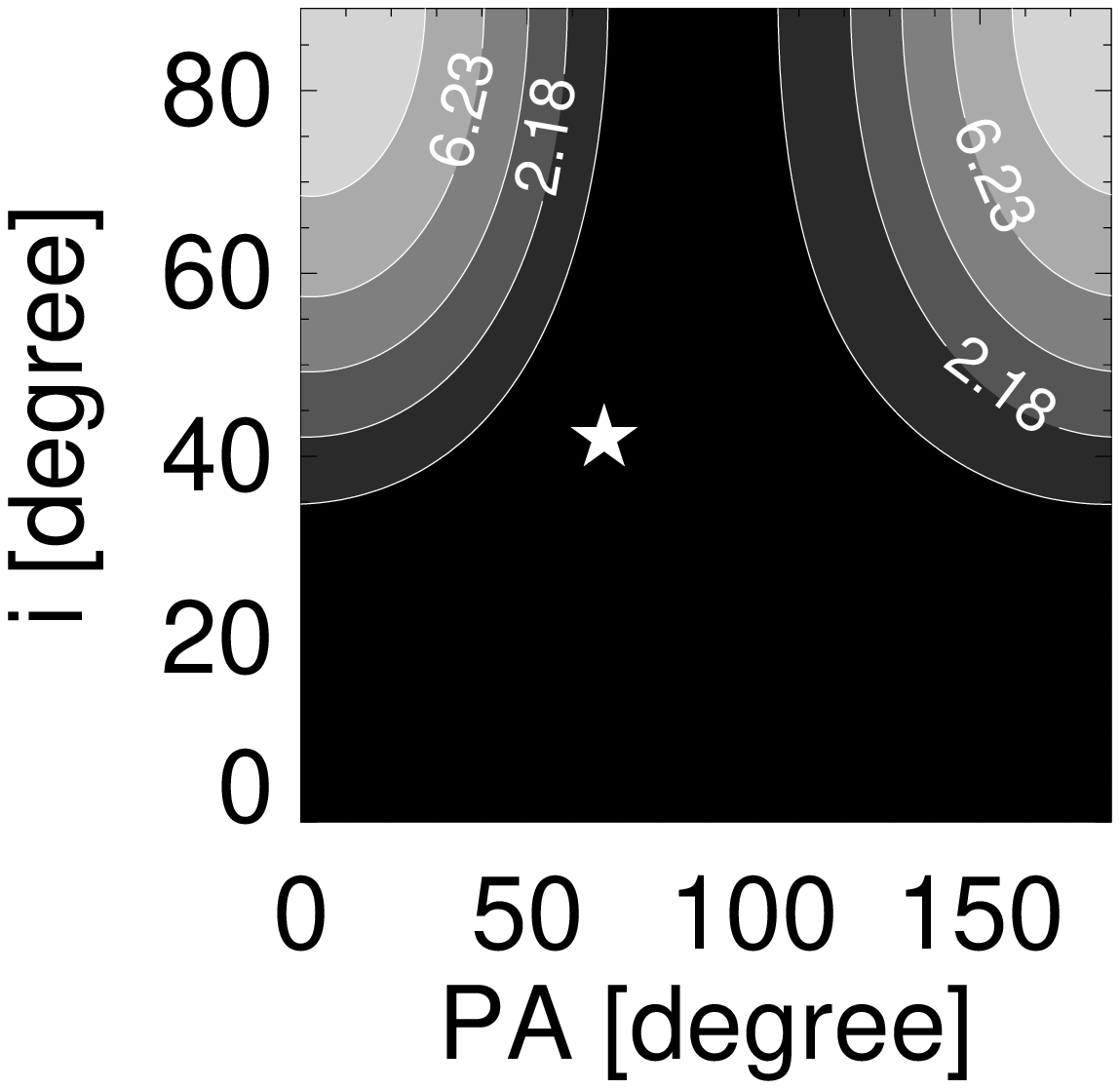}
\caption{{\it Top panels}: reduced $\chi^2_{\mathrm{r}}$ maps for the binary
  model. {\it Bottom panels}: reduced $\chi^2_{\mathrm{r}}$ maps for the uniform
  ring model. The star symbol corresponds to values found for the best
  fit with the lowest reduced $\chi^2_{\mathrm{r}}$.\label{fig:bin}}
\end{center}
\end{figure}

\end{document}